# Challenges on Optical Printing of Colloidal Nanoparticles

*Ianina L. Violi, [1,2*] Luciana P. Martinez, [1] Mariano Barella, [1] Cecilia Zaza, [1,3] Lukáš Chvátal, [4] Pavel Zemánek, [4] Marina V. Gutiérrez, [5] María Y. Paredes, [5] Alberto F. Scarpettini, [5] Jorge Olmos-Trigo, [6] Valeria R. Pais, [3] Iván Díaz Nóblega, [3] Emiliano Cortes, [7] Juan José Sáenz, [6] Andrea V. Bragas, [3] Julian Gargiulo, [1,7*] and Fernando D. Stefani [1,3*]*

1. Centro de Investigaciones en Bionanociencias (CIBION), Consejo Nacional de Investigaciones Científicas y Técnicas (CONICET), Godoy Cruz 2390, CABA, Argentina

2. Instituto de Nanosistemas, UNSAM-CONICET, Av. 25 de Mayo 1021, San Martín 1650, Argentina.

3. Departamento de Física, Facultad de Ciencias Exactas y Naturales, Universidad de Buenos Aires, Güiraldes 2620, CABA, Argentina

4. Institute of Scientific Instruments of the Czech Academy of Sciences, v.v.i., Czech Academy of Sciences, Královopolská 147, 612 64 Brno, Czech Republic

5. Grupo de Fotónica Aplicada, Facultad Regional Delta, Universidad Tecnológica Nacional, 2804 Campana, Argentina

6. Donostia International Physics Center (DIPC), Donostia-San Sebastián, País Vasco, España

7. Chair in Hybrid Nanosystems, Nanoinstitute Munich, Faculty of Physics, Ludwig-Maximilians-Universität München, 80799 München, Germany

ABSTRACT

While colloidal chemistry provides ways to obtain a great variety of nanoparticles, with different shapes, sizes, material composition, and surface functions, their controlled deposition and combination on arbitrary positions of substrates remains a considerable challenge. Over the last ten years, optical printing arose as a versatile method to achieve this purpose for different kinds of nanoparticles. In this article we review the state of the art of optical printing of single nanoparticles, and discuss its strengths, limitations, and future perspectives, by focusing on four main challenges: printing accuracy, resolution, selectivity, and nanoparticles photostability.

Colloidal chemistry allows the preparation of nanoparticles (NPs) with different compositions, sizes, and morphologies, which in turn define their unique physical and chemical properties that are impossible to obtain in bulk materials. In order to study them at a single particle level and to take advantage of their features in nanofabrication of circuits and devices, it is necessary to develop methods to bring the colloidal NPs from the liquid phase to specific locations of a solid substrates. However, obtaining arbitrary arrays of colloidal NPs with different compositions, morphologies, and functions is still an open challenge.[1]

The simplest approaches to pattern surfaces with colloidal NPs are self-assembly methods.[2] In these strategies, NPs self-organize into the most thermodynamically stable configuration as a result of the interplay of potentials taking part in the assembly process. Large-patterned areas can be obtained, but with limited flexibility of pattern design. In template-assisted self-assembly, the substrate is chemically or physically patterned before the assembly using top-down lithographic methods.[2–4] This way, the deposition of particles only takes place in previously selected positions. Geometric flexibility is achieved at the cost of lengthening and complicating the fabrication procedure. Still, arbitrary combination of different types of particles remains impossible and changing the pattern requires fabricating new templates.

There is a family of techniques that provide higher flexibility of pattern design by using light to dynamically create regions of the substrate with higher affinity for the colloids. Optoelectronic Tweezers use light and an electric bias to sculpt a potential landscape on a photosensitive substrate.[5,6] The resulting non-uniform electric field exerts dielectrophoretic forces on NPs that push them towards the illuminated regions. In opto-thermophoretic printing,[7] absorbing substrates are illuminated to generate a thermal field that drives the NPs towards the hot illuminated regions by means of thermophoretic forces. Bubble-pen lithography[8,9] employs a laser beam to generate microbubbles at the interface of a colloidal suspension and a plasmonic substrate. The bubbles capture and immobilize the NPs. All these techniques allow the serial deposition of individual NPs with high accuracy and versatility of pattern design. However, they require specialized substrates such as highly absorbing or conducting films.

By contrast, optical printing uses optical forces to capture individual NPs from the colloidal suspension, push them towards the substrate, and fix them into predefined locations through van der Waals attraction. This technique is especially suitable for metallic or high-index dielectric NPs, because their plasmon or Mie-like resonances strongly enhance the optical forces, that are otherwise low for sub-micrometer particles.[10–12] Optical printing enables the selective immobilization of different kinds of NPs using light of different wavelengths, as long as the NPs present spectrally differentiable resonances.[13,14]

In 2010, the groups of Norbert Scherer at the University of Chicago[15] and Jochen Feldmann at the Ludwig-Maximilians-Universität of Munich[16] demonstrated almost simultaneously the optical printing of single spherical Au NPs. The experiments presented by the Scherer group consisted in two steps. First, a single Au NP (40 nm in diameter) was trapped with an optical tweezer working at 817 nm. Then, by manually moving the focus, the optically trapped NP was pushed against the substrate and finally attached to it by van der Waals forces. In the work by the group of Feldmann, single Au NPs (80 nm in diameter) were captured from the colloidal suspension and pushed towards the substrate, without any previous stable trapping step, using a focused laser beam tuned to the plasmon resonance of the NPs. Since then, optical printing has been employed to produce patterns of single[13,14,17–21] or aggregated[22–24] NPs of different sizes, shapes, and materials, including gold, silver, and silicon NPs, onto various substrates.

Here, we review and discuss the advances made to date on optical printing of colloidal NPs, organized in four main challenges: printing accuracy, spatial resolution, selectivity, and photostability of NPs (Figure 1). First, we briefly describe the fundamentals of the method. Then, the advances made with regard the four challenges are evaluated, including in some cases unpublished data from our group. Finally, outlook and perspectives on the field are presented.

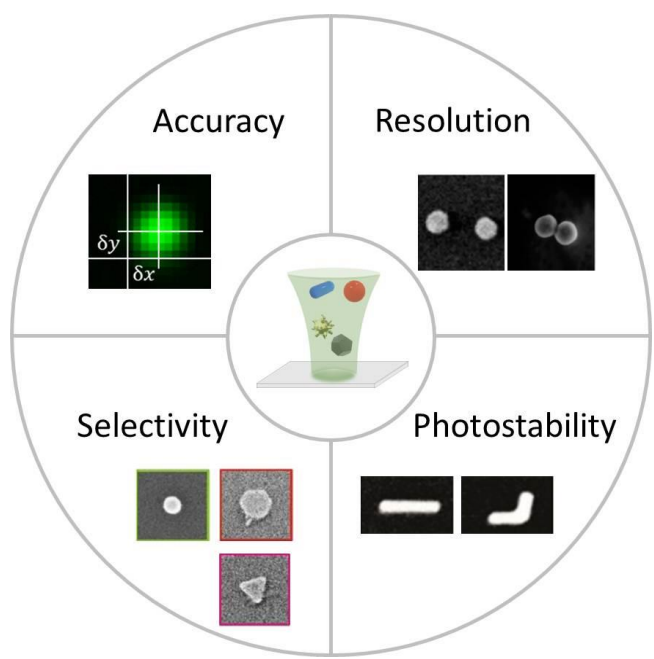

**Figure 1.** Advances in optical printing are discussed in the context of four challenges: Accuracy (figure adapted with permission from Ref. 18. Copyright 2017 American Chemical Society), Resolution (figure adapted with permission from Ref. 19. Copyright 2017 American Chemical Society), selectivity (figure adapted with permission from Ref. 14, Copyright 2016 American Chemical Society), and photostability (figure adapted with permission from Ref. [25]. Copyright 2016 American Chemical Society).

*Fundamentals of optical printing*

For optical printing, a colloidal suspension of NPs with certain net electric charge is put in contact with a substrate with the same charge sign. In this way, spontaneous deposition of the NPs onto the substrate is avoided by electrostatic repulsion. The basis of optical printing is to overcome the repulsion barrier on specific locations of the substrate by applying optical forces on the NPs. This is achieved using a laser beam tightly focused onto the substrate.

The process of optical printing can be summarized in three steps as schematized in Figure 2a.

1) A NP performs free diffusive Brownian motion until it randomly reaches the beam focus. The duration of this diffusive motion depends on NP concentration and laser power, and typically lies in the order of a few seconds. The region of space where optical forces dominate over Brownian motion is defined as the "capture volume", depicted with a green dashed line in Figure 2a.[11] The capture volume can be estimated as the region where the Peclet number is higher than 1, as described in Gargiulo *et al.*[18]

2) Once in the capture volume, optical forces drive the NP towards the substrate. This motion is overdamped (Reynolds number is <<1) and optical force field lines can be interpreted as velocity lines or trajectories. The time scale for this guided motion inside the capture volume is on the order of milliseconds.

3) An interaction between the NP and the substrate appears at separations of a few tens of nanometers, consisting of an electrostatic repulsion $F_{el}$ and an attractive van der Waals force $F_{vdW}$. The total interaction $F_{DLVO} = F_{el} + F_{vdW}$ may be obtained by differentiation of the interaction energies between a charged sphere and a charged surface using the Derjaguin-Landau-Verwey-Overbeek (DLVO) formalism [26]. Figure 2b (top) shows an example of DLVO potentials for a 60 nm Au NP near a flat charged surface. As the NP approaches the substrate, the electrostatic repulsion generates a barrier that prevents the NPs from reaching the substrate. There is a minimum axial force (threshold force, $F_{th}$) that must be applied to the NP to overcome the repulsion barrier. If the NP is pushed over the repulsion barrier, it gets fixed on the substrate by the attractive van der Waals force. It is then said that the NP has been printed.

Figure 2b (bottom) shows a comparison of $F_{DLVO}$ for the cases of a Au NP and a Ag NP, both 60 nm in diameter, close to a glass substrate. In this case, the difference in the threshold force ($F_{th}$) relies on the different zeta potential (surface charge) of the two types of NPs, as the rest of the parameters are identical.

*Implementation and examples of optical printing*

Figure 2c shows schematically an example setup for optical printing. It basically consists of an upright microscope where the printing laser beam is focused onto the sample by the microscope objective. Usually, a water immersion objective is used in direct contact with the colloidal suspension. In fact, to protect the objective from the adhesion of NPs or other colloid components, a transparent film (food cling film, for instance) may be used to protect the objective lens. Alternatively, the substrate and the colloid may be set in a (flow) chamber, enabling the use of other types of objectives. Various techniques or detection channels may be included to monitor the printing process or characterize the NPs. For example, dark-field imaging is usually employed to monitor the printing process and to visualize the produced patterns of NPs.[16] Confocal detection of the scattering by the NPs provides a useful signal to automate the printing process.[17] Scattering, photoluminescence, and/or Raman spectroscopy are suited for characterizing the printed NPs.[20,27]

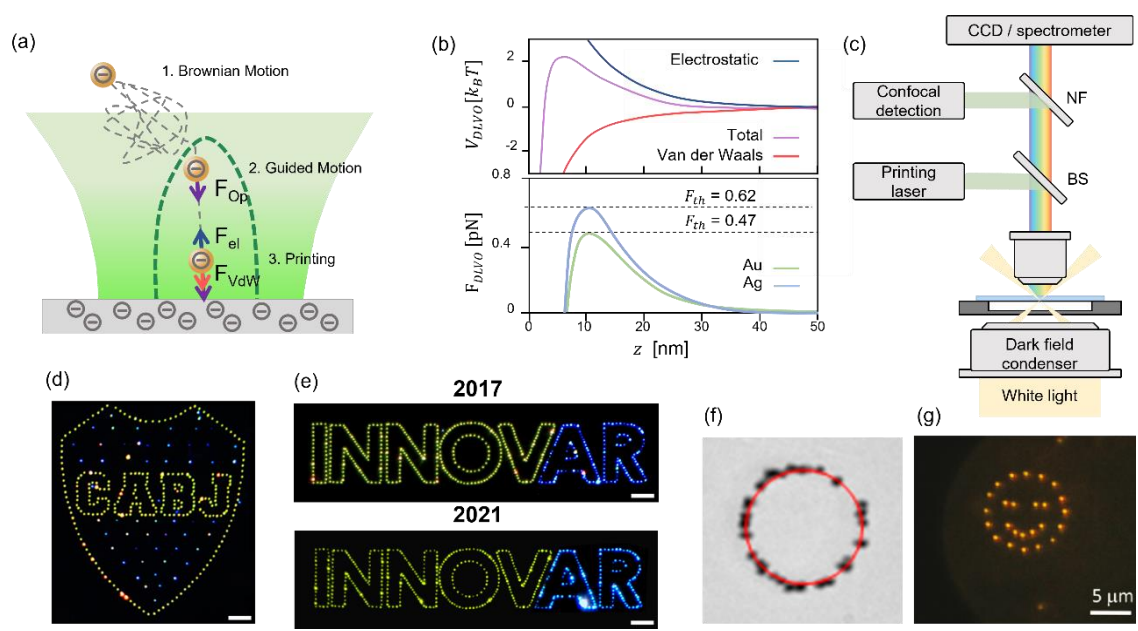

**Figure 2:** (a) Schematic of the optical printing process (b) Calculated DLVO potentials (top, 60 nm Au NP) and force (bottom, 60 nm Au and Ag NPs) of a NP versus the distance to a flat charged surface. Maximum repulsive forces (DLVO threshold barriers, $F_{th}$) are indicated with horizontal dashed lines. (c) Scheme of an optical printing setup. BS: Beam splitter; NF: Notch filter. (d, e) Exemplary dark-field images array of Au (green) and Ag (blue) NPs (60 nm in diameter) fabricated by optical printing. Scale bars: 10 μm. (f) 200 nm Ag NPs deposited by optical trapping with a ring pattern laser trap. The red circle is 6 μm in diameter. (g) 80 nm Au NPs deposited by optical force stamping lithography, employing a spatial light modulator to produce a "smiley" pattern. (b) Adapted from Gargiulo *et al.*[18] Copyright 2017 American Chemical Society. (f)

Reproduced with permission from Bao et al.[28] Copyright 2014 American Chemical Society. (g) Reproduced with permission from Ref. [29] Copyright 2011 American Chemical Society.

Arrays of NPs can be fabricated by sequentially printing single NPs on the desired positions. This is achieved by using a sample or a beam scanner to move the printing laser to the next position once a NP is printed. Figure 2d and 2e show dark-field images of two arrays made by optical printing Au (green) and Ag (blue) NPs, both 60 nm in diameter. A remarkable property of optically printed NP arrays is their high stability over time. For example, the pattern shown in Figure 2e was fabricated in 2017. A recent inspection shows no significant changes after 4 years. All NPs remained in their original positions. In between, the sample was dried, manipulated and stored in a drawer at room temperature. In addition, Section S1 of the electronic supplementary information (ESI) shows images of samples that were subjected to vacuum and electron microscopy imaging, also showing high stability.

Alternatively, arrays of NPs may be fabricated by parallel optical printing. Using a spatial light modulator or phase masks, extended foci may be generated to produce optical printing over lines, rings (Figure 2f) or any other shape.[28] Also, multi-beam arrays may be used to print several NPs in parallel[29] (Figure 2g).

*Challenge 1: Accuracy of optical printing*

The first challenge is to immobilize the NPs with the maximum possible positional accuracy. Despite being an optical method, already in the first experiments it became clear that the positional accuracy of optical printing was not restricted by the diffraction limit of light. In 2010, Urban et al.[16] printed 80 nm Au NPs with a laser of 532 nm, a wavelength near to the plasmon resonance of the NPs. They printed NPs along a line and estimated the printing precision as the average deviation from the line, which was found to be 50 nm (Figure 3a). They also found that the best accuracy was achieved by using the minimum laser power needed to overcome the electrostatic barrier. In addition, the influence of the electrostatic barrier strength was studied by using substrates coated with increasing numbers of polyelectrolyte layers. It was found that higher laser powers were required for printing when more layers were present, indicating a stronger repulsion.

Larger deviations from the target line were observed for the higher number of layers, obtaining an accuracy of ~100-150 nm.

In the second seminal work of optical printing, Guffey et al.[15] used light out of resonance (λ = 817 nm) to first trap 40 nm Au NPs, and then bring them into contact with a glass substrate by manually moving the focus. The printing accuracy was found to be around ~100 nm, which was larger than the standard deviation of the particle position fluctuations within the trap (~30 nm). Contrary to Urban et al., they reported that the best accuracy was obtained for higher powers. Hence, these first two works demonstrated sub-diffraction printing precision and identified the laser power and the electrostatic barrier as factors influencing the positional accuracy. Still, the physical mechanisms limiting the printing precision were unclear, particularly in view of the contradicting dependencies on the laser power.

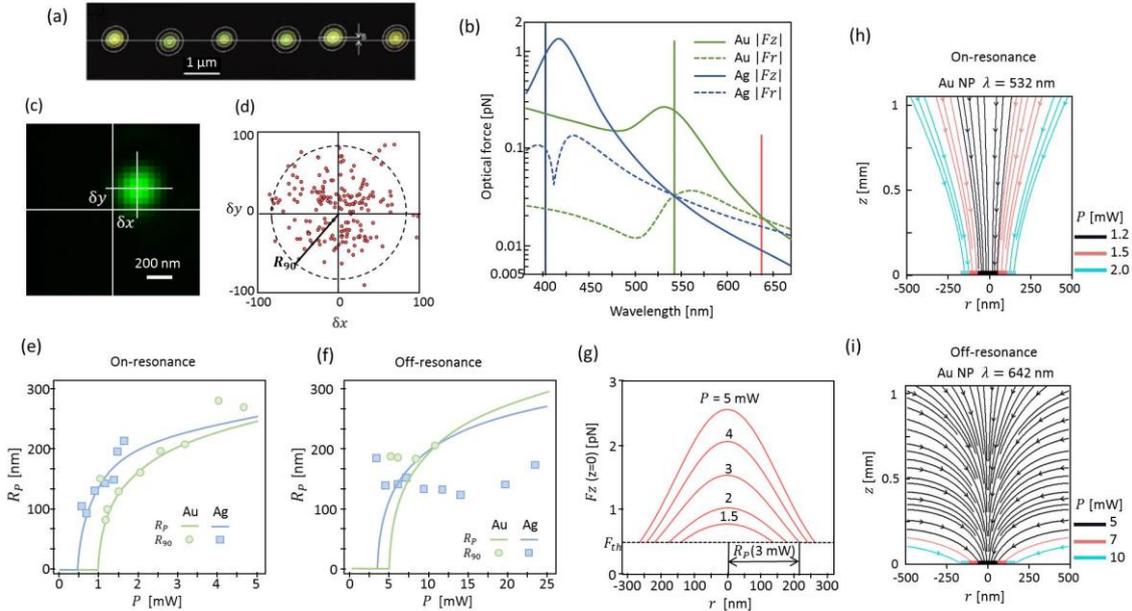

**Figure 3.** (a) Dark-field image of a line of printed Au NPs used to estimate printing accuracy. White line is the target line, including the results of the 2D Gaussian fit of each particle used to determine NPs position. (b) Absolute value of the maximum axial ($F_z$) and radial ($F_r$) forces versus wavelength for 60 nm Ag and Au NPs placed at the center of a diffraction limited Gaussian focus with NA = 1.0, and incident laser power of P = 1 mW. The three vertical dashed lines correspond to the printing wavelengths used to print Ag and Au NPs on- and off-resonance (405 nm, 532 nm and 640 nm. (c) Example of a scattering confocal image used to determine the center of the printed NP and the printing error δx and δy. (d) 2D scatter plot of the printing error of 196 printed Au NPs. The circle (of radius $R_{90}$) delimits the region where 90% of the NPs were printed.

(e, f) Experimental values of $R_{90}$ as a function of laser power $P$ for 60 nm Au (circles) and Ag (squares) NPs printed on-resonance (e) and off-resonance (f). Solid lines: $R_p$ as a function of laser power $P$. (g) Calculated optical axial force $F_z$ at the substrate (focal plane, $z = 0$) as a function of the radial position, for Au NPs at 532 nm and different beam powers. The horizontal line indicates the threshold for optical printing imposed by the repulsive DLVO barrier ($F_{th}$). The region $F_z > F_{th}$ delimited by a printing radius $R_p$, schematically shown for laser power P = 3 mW. (h-i) Calculated trajectories for 60 nm Au NPs at different powers and wavelengths. Only the trajectories that end inside the printing radius $R_p$ are shown. Pink and light blue lines are those trajectories that become available when increasing the power with respect to the lower value. Colored boxes at $z = 0$ depict the printing radius $R_p$ at each power. (a) Reproduced with permission from Urban et al.[16] Copyright 2010 American Chemical Society. (b-i) Adapted with permission from Gargiulo et al.[18] Copyright 2017 American Chemical Society.

Later, Gargiulo et al.[18] performed a systematic study of the printing accuracy for 60 nm Au and Ag NPs at different printing conditions. They characterized the printing accuracy versus the laser power for wavelengths on-resonance and off-resonance with respect to the plasmon resonance of the NPs. For that purpose, they performed calculations and measurements using three wavelengths: 405 nm (focused beam waist of 226 nm), 532 nm (focused beam waist of 265 nm), and 640 nm (focused beam waist of 319 nm).

Figure 3b shows, for 60 nm Au and Ag NPs, the magnitude of radial ($F_r$) and axial ($F_z$) optical forces versus wavelength, at the center of a diffraction limited Gaussian focus. For both materials, on-resonance (laser wavelength: 532 nm for Au NP and 405 nm for Ag NP) axial forces are approximately 10-fold stronger than radial forces. In contrast, off-resonance (laser wavelength: 640 nm for Au NP and 532 nm for Ag NP), both axial and radial forces have comparable magnitudes.

For each condition, at least 200 individual NPs were printed and localized using confocal scattering images, as shown in Figure 3c. The difference between the target position and the NP position was defined as the printing error (δx, δy). An exemplary scatter plot of the printing errors obtained for a given set of conditions is shown in Figure 3d. Interestingly, although the experiments were carried out with linearly polarized light, the printing error showed circular symmetry independently of the many different printing conditions used, indicating that the laser polarization does not influence the printing precision. To quantify the printing accuracy, the authors used the printing radius $R_{90}$, defined as the radius containing 90% of the printed NPs (dotted line in the scatter plot of Figure 3d).

Figures 3e and 3f show the experimental printing radius $R_{90}$ versus the laser power determined for Au and Ag NPs when they were optically printed on- and off-resonance, respectively. On-resonance, the Au NPs were printed at 532 nm (green circles) and the Ag NPs were printed at 405 nm (blue squares). For both, the precision decreases when the laser power increases. Off-resonance, the Au NPs were printed at 640 nm (green circles) and the Ag NPs at 532 nm (blue squares). Interestingly, in this case the $R_{90}$ was found to be independent of the laser power and larger than the best achieved precision in the resonant case.

As expected from the balance between the optical axial force and the DLVO interaction with the substrate, it was not possible to print particles below a certain threshold of laser power $P_{th}$ ($F_{th}$, Figure 2b). Calculating the optical and the DLVO forces enabled a quantitative prediction of this behaviour.[18] In Figure 3g, $F_z$ at the substrate is compared to $F_{th}$. The region where optical printing is possible is determined by the condition $F_z > F_{th}$ which defines a maximum radius depending on the laser power. This maximum radius was called the theoretical "printing radius" $R_P$. In Figure 3e, $R_P$ is shown along with $R_{90}$ as a function of the laser power for resonant printing of Ag NPs (blue solid line) and Au NPs (green solid line). The theoretical $R_P$ represents very well the experimental values $R_{90}$ and correctly predicts the value for the threshold power $P_{th}$, remarkable facts given that there were no free parameters involved in the calculations. On the other hand, for the off-resonance case, it can be seen in Figure 3f that the printing radius $R_P$ does not predict the nearly constant experimental $R_{90}$, indicating that a deeper analysis is necessary.

A more detailed discussion is feasible by considering the possible trajectories imposed by the optical forces on the NPs to be printed. Figures 3h and 3i show the trajectories for NPs optically printed on-resonance and off-resonance, respectively, for three laser powers. The achieved $R_p$ with each laser power is depicted by colored boxes at $z = 0$ (*i.e.,* at the substrate position). On-resonance, increasing the laser power leads to more trajectories becoming available for printing that enlarge $R_p$ (*i.e.,* lower precision). By contrast, off-resonance, most trajectories reaching the substrate become available as soon as the laser power surpasses the threshold for printing. Further increments of the laser power only lead to minor or negligible increments of the number of trajectories available for printing, explaining the independence of $R_{90}$ on laser power for off-resonance printing.

Overall, it can be concluded that the best accuracy for optical printing spherical NPs with Gaussian beams is achieved on resonance and close to threshold power $P_{th}$. At these conditions, 90% of the NPs are printed within 100 nm of the target, and the radial root-mean-square deviation is in the order of 50 nm. These numbers are considerably smaller than the waist of the printing beam and are comparable to the diameter of the NPs. The prediction of the printing accuracy requires two steps: 1) calculating the printing radius from the balance between the optical force and the repulsion by the substrate, and 2) an analysis of the possible trajectories to determine the regions of the printing radius that are likely to be filled.

We note that the above analysis is valid for printing freely diffusing colloidal NPs, where each NP may enter the capture volume at different positions. A higher precision could be attained if the NPs were forced into the capture volume from the same position, thus reaching the substrate following a smaller set of trajectories. This could be achieved, for example, if the NPs were loaded into the printing beam using an additional optical tweezer, with a repeatable entry position. Another possibility would be to use a microfluidic chamber, where a fluid flow would drive the NPs towards the focus with controlled direction and speed.

Until now, mostly Gaussian beams have been explored. Engineering more complex optical profiles leading to smaller printing radius may also open the door to higher attainable precisions. For example, an additional counter-propagating doughnut-shaped beam could further confine the printing region, increasing the final precision.

*Angular accuracy*

In the case of printing non-spherical NPs, the additional challenge of controlling their orientation arises. In 2012, Ling *et al.*[30] studied the printing of Au nanorods (NRs) of (25×73 ± 10) nm in size and longitudinal plasmon resonances around 700 nm, on a glass substrate with a thick film of polyvinyl alcohol. Using a linearly polarized, off-resonance laser at 1064 nm, they trapped and moved single Au NRs towards the surface. With this strategy, they could optically print Au NRs with preferential orientation parallel to the polarization of the trapping beam. Figure 4a shows dark-field images of the Au NRs printed with two orthogonal polarizations. The red arrows show the direction of the polarization used to print each NR. The light blue arrows represent the polarization of the white lamp used to acquire the dark-field image. The scattering intensity of the NRs is stronger when the illumination polarization is parallel to the polarization of the trapping beam, indicating that the Au NRs were printed with their long axis parallel to the polarization of the trapping beam.

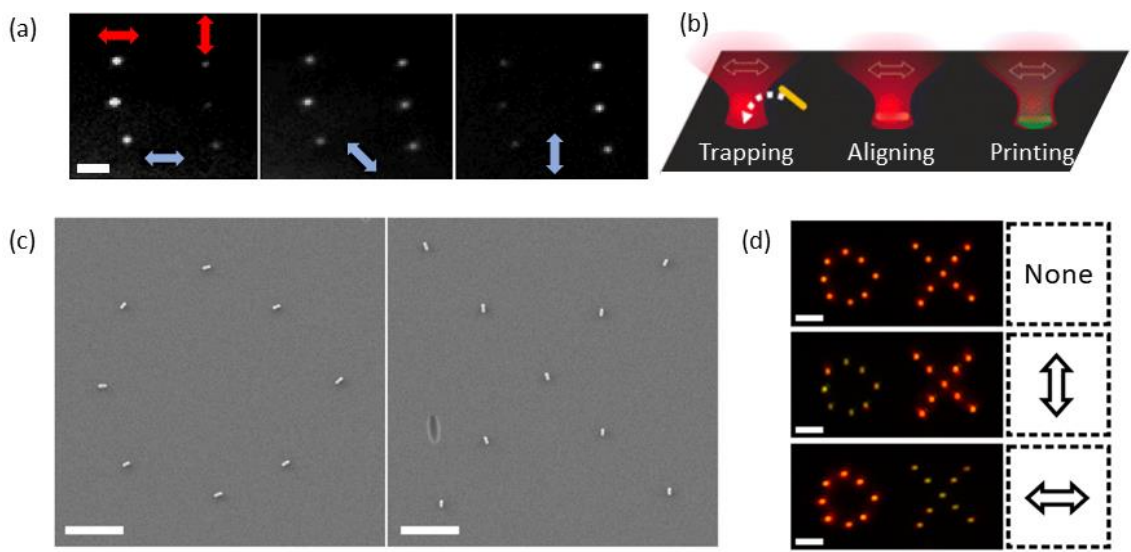

**Figure 4.** (a) Polarized dark-field images of Au NRs fixed by optical trapping and printing with different polarizations. The red arrows indicate the polarization of the trapping and printing laser (1064 nm wavelength). The light blue arrows indicate the polarization direction of the white light used for dark-field imaging. Scale bar: 2 µm. (b) Dual-color experimental scheme by Do *et al.* used to trap (1064 nm, depicted as a red beam) and print (532 nm, green) Au NRs. (c) SEM Images of an "OX" pattern of printed Au NRs. Scale bar: 1 µm. (d) Polarized dark-field images of the "OX" pattern. Scale bar: 2 µm. The direction of the linear polarizer in front of the detector is indicated, showing opposite anisotropic response of the "O" and "X" letters. (a) Adapted from reference Ling *et al.*[30] Copyright IOP Publishing. Reproduced with permission. All rights reserved. (b-d) Adapted with permission from Do *et al.*[21] Copyright 2013 American Chemical Society.

Further control on the orientation Au NRs was achieved in 2013 by Do *et al.*[21] using a two-step process of trapping and subsequent printing. First, a linearly polarized, 1064 nm focused laser beam was used to trap and orient NRs with longitudinal plasmon resonance at 724 nm. Then, a co-aligned linearly polarized 532 nm laser was switched on, to print the NR on the substrate. The process is schematized in Figure 4b. The authors found that the highest orientational control was achieved if trapping and printing lasers were polarized perpendicularly to each other. Such configuration reduces the torque over the pre-aligned nanorod as the trapping and printing laser are tuned to match the longitudinal and transversal mode, respectively. Figure 4c shows Scanning Electron Microscopy (SEM) images of two printed patterns, fabricated using two orthogonal polarizations of the trapping beam. Clearly, the Au NRs were printed with a preferential orientation parallel to the polarization of the trapping beam. The dark-field images shown in Figure 4d further confirm this result. Illumination was not polarized. In the upper panel, no polarizer was included the detection path, and both patterns show similar colors. When the polarization detector is parallel to the polarization used for trapping, the corresponding pattern appears brighter and in red color as the scattered light comes predominantly from the longitudinal resonance. Also, the interplay between the trap stiffness, proportional to the 1064 nm laser power, and the scattering force, proportional to the 532 nm printing laser power, was found to be essential.

In summary, it can be concluded that NPs with anisotropic optical properties can be optically printed with preferential orientations using linearly polarized laser beams.

*Challenge 2: Lateral resolution*

Another important challenge of optical printing concerns its resolution, understood as the minimum distance at which two particles can be printed in a controlled manner. Already in 2013, a limitation to print nanoparticles at short distances was identified. Urban *et al.*[31,32] observed that *"printing of two or more gold nanoparticles in close proximity to each other has been a challenge and separation distances below 300 nm have not been achieved experimentally."* A strategy to overcome this problem was presented by Urban *et al.*[31] and is depicted in Figure 5a. Dimers and trimers of 80 nm Au NPs originally printed on glass were transferred to a thermo-responsive polymer film. Upon heating, the polymer shrunk, decreasing by a factor of 2.3 the average interparticle distances. The minimum achieved separation distance between NP centers was 120 nm, and 20 to 40 nm between NP surfaces.

In 2014, Bao *et al.* systematically studied the lateral resolution.[28] They used an optical line trap[28] at a wavelength of 800 nm to print chain patterns of 200 nm Ag NPs, as shown in Figure 5b. They attempted to print pairs of chains at different separation distances $\Delta x$ and identified three regimes (Figure 5b): i) A *non-interacting* regime (green shaded) for set distances $\Delta x > 600$ nm where the chains of particles can be printed with the desired separation; ii) An *interacting* regime (yellow) for 300 nm $< \Delta x <$ 600 nm, characterized by deviations of the printed NPs toward larger separation distances than the set distance; and iii) An *unstable* regime (pink) for $\Delta x < 300$ nm, resulting in the loss of control of the printing process where some of the incoming particles are printed onto the existing chain, forming aggregates.

In 2016, Gargiulo *et al.*[17] performed another study on the lateral resolution, this time manipulating single NPs. They printed pairs of individual 60 nm Ag and Au NPs and compared the set separation distances to the experimentally achieved center to center interparticle distance. NPs were printed on-resonance, using wavelengths of 405 nm and 532 nm for Ag and Au NPs, respectively. Extinction spectra of the NPs are shown in Section S2 of the ESI. For pairs of Au NPs, three different regimes were also observed, as shown in Figure 5c (circles). For set distances above 350 nm, NPs were correctly printed (*non-interacting regime*). For set distances between 250 nm and 350 nm, NPs were printed at separation distances larger than the set distance (*interacting with printing regime*). For set distances smaller than 250 nm, it was impossible to print the second NP to form the dimer (*interacting without printing regime*). These results strongly suggested the presence of a light-induced repulsive force that sets in as soon as the already printed NP is illuminated. To further investigate this repulsion, Ag-Au heterodimers were fabricated at different set distances. In this case, the results depended on the printing order (Figure 5c, squares). If the Au NP is printed first, the *non-interacting regime* extends down to separation distances of 250 nm, and it is always possible to print the second (Ag) NP, showing that the repulsion is weaker than in the Au-Au case. On the other hand, if the Ag NP is printed first, the repulsion disappears completely, and optical printing is possible and free of interactions for every set distance. Figure 5d shows dark-field and SEM images of exemplary Ag-Au heterodimers, printed with a set distance of 60 nm, equal to the NP diameter. Remarkably, it was possible for the first time to connect a pair of NPs using optical printing. Furthermore, the orientation of the dimer was well controlled.

The experiments described so far confirm that the resolution of optical printing may be limited by a light-induced repulsion. First, the transitions between *non-interacting* to *interacting regimes* occur at set distances comparable to the laser beam waist at the focal plane (~250 nm), *i.e.*, where the already printed NP is illuminated. Second, the magnitude of the repulsion depends on the optical interaction between the first NP and the printing wavelength of the second one. It is strong for Au at 532 nm, weak for Au at 405 nm, and negligible for Ag at 532 nm, consistent with the optical extinctions at each respective wavelength (see Figure S2). If the first printed NP is transparent to the printing wavelength of the second NP, the repulsion is avoided, and optical printing can place NPs at arbitrarily small separations.

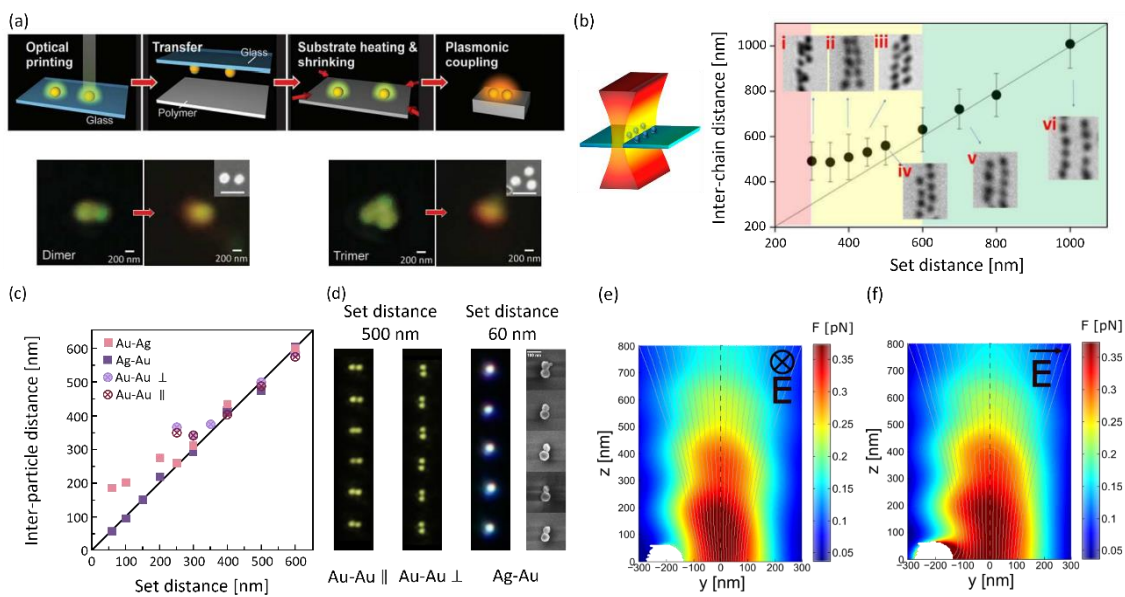

**Figure 5.** (a) Top, illustration of the process of transferring optically printed Au NPs to a thermo-responsive polymer sheet. Upon heating, the polymer shrinks, decreasing the separation between printed NPs. Bottom, example dark-field and SEM images of a printed dimer and trimer of Au NPs before and after heating. (b) Left: schematic of a line optical trap used for printing chains of NPs. Right: experimentally observed inter-chain mean separation distance as function of set separation. (c) Measured inter-particle distances vs. set distance for different optically printed NP dimers. (d) Dark-field and SEM images of optically printed Au-Au and Au-Ag NP dimers. (e-f) Optical force maps produced by a Gaussian beam on a 60 nm Au NP in the presence of a fixed NP 200 nm away from the beam center. The polarization of the beam is perpendicular (e) and parallel (f) to the y axis. White areas excluded from the map correspond to inter-penetrated NPs. Grey lines are force field lines. The modulus of the force is color coded. (a) Reproduced with permission from Ref. [31]. Copyright 2013 WILEY-VCH Verlag GmbH & Co. KGaA, Weinheim. (b) Reproduced with permission from Bao *et al*.[28] Copyright 2014 American Chemical Society. (c-d) Adapted with permission from Gargiulo *et al*.[17] Copyright 2016 American Chemical Society.

The next step was to determine the nature of the light-induced repulsion, whether it was due to scattering, absorption, or both. Scattering by the already printed NP could, in principle, distort the laser focus shifting the position of maximum axial force away from the target printing position. In addition, during the optical printing process, both the fixed and the approaching NP polarize, with induced dipoles parallel to the electric field. Interaction between the induced dipoles would generate additional forces in a similar fashion as it happens in optical binding.[33–35] So, if the repulsion is due to scattering, it is sensible to expect a dependence with the polarization of light. This was tested experimentally by printing Au-Au dimers with a linearly polarized 532 nm laser at different set separation distances with two orientations: parallel and perpendicular to the polarization. The results are shown in Figure 5c, and exemplary dark-field images are shown in Figure 5d. Interestingly, there are no distinguishable differences between the two orientations. Furthermore, calculations of the total optical force field in the presence of a fixed NP cannot explain the observed repulsion. For example, Figures 5e and 5f show the trajectories capable of printing a 60 nm Au NP when a linearly polarized laser beam is placed 200 nm away from an already printed NP. The plot in Figure 5e corresponds to the laser polarization perpendicular to the dimer axis, while Figure 5f for the parallel polarization. The distance of 200 nm corresponds to a configuration where it is impossible to print a second NP (see Section S3 of ESI for details on the calculations). However, the calculations show many field lines that end in the substrate for both polarizations. Furthermore, for the parallel polarization (Figure 5f) there are even trajectories that end closer to the fixed NP.

While scattering and optical binding may generate considerable optical forces in other configurations, they were not the main cause of the observed light-induced repulsion, which was then ascribed to absorption by the printed NP. When a laser focus is set to print a NP near an already printed NP, both NPs absorb light and generate a local temperature increase. In 2017, Gargiulo *et al.*[19] studied the influence of photothermal heating in the printing process by implementing different strategies to enhance heat dissipation. First, they printed Au-Au 60 nm dimers with different set gaps on a sapphire substrate, a material with higher thermal conductivity ($\kappa \sim 20$ W/m K) than glass ($\kappa \sim 1$ W/m K). Figures 6a and 6b show the calculated temperature increase for a 60 nm Au NP on a glass and a sapphire substrate, respectively. In both cases, the NP is immersed in water and at the center of Gaussian laser beam of 532 nm wavelength, 266 nm waist, and 1.2 mW of power. On sapphire, the temperature increase is 7-fold smaller than on glass. Figure 6c shows the measured gaps of dimers of 60 nm Au NPs optically printed on glass and sapphire, versus the set gap. Remarkably, on sapphire, the light induced repulsion is considerably reduced, and optical printing of the second NP is possible at any set gap. An *interacting with printing regime* for set gaps smaller than 200 nm is observed, leading to experimental gaps as small as 150 nm. On glass, optical printing became impossible at set gaps smaller than 230 nm (horizontal dashed line in Figure 6c).

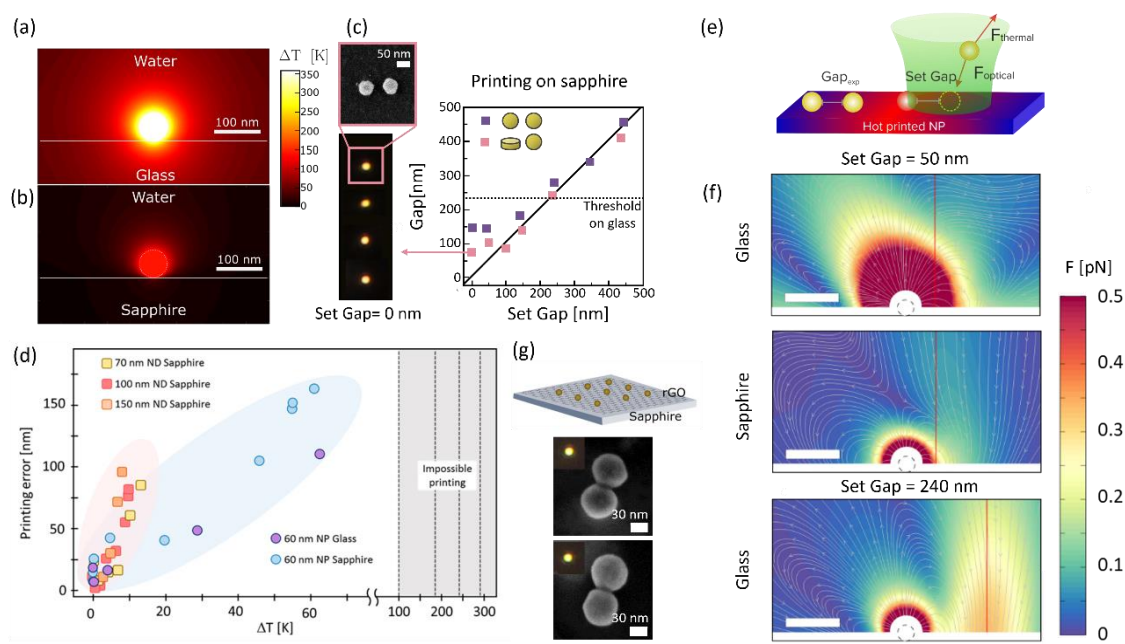

**Figure 6.** (a-b) Temperature increase map of a single Au 60 nm sphere in the center of a Gaussian beam on (a) glass and (b) sapphire substrates. (c) Measured experimental gap of Au NP-NP and ND-NP dimers for different set gaps between the surface of the structures. The horizontal dashed line shows the minimum attainable gap for two Au NPs in glass. The inset shows exemplary dark-field and SEM images of ND-ND dimers fabricated with a set gap of 0 nm. (d) Mean printing error of the second particle in the dimer versus the temperature increase of the first particle. Vertical dashed lines in the grey area indicate experimental conditions where it was impossible to print. (e) Schematic of the Au−Au dimers printing experiment and its relevant parameters and forces. As the laser beam illuminates the NPs, the local temperature increases. (f) Maps of the net force (optical + thermophoresis) acting on a 60 nm Au NP for positions around a fixed NP at (0,0) at different substrates and beam positions. Streamlines indicate the direction of the net force, and its magnitude is color coded. Scale bars: 200 nm. (g) Top: Schematic of an optically printed grid of Au NPs on a sapphire substrate coated with reduced graphene oxide. Bottom: dark-field images of 60 nm Au NPs optically printed on rGO. Figure adapted with permissions from Gargiulo *et al.*[19] Copyright 2017 American Chemical Society.

Then, optical printing a Au NP next to a Au nanodisk (ND) was tested. Disks were fabricated on sapphire substrates using top-down lithography. Since they have a larger contact area with the substrate, a better heat dissipation is expected. Figure 6c (pink squares) shows the measured gap versus the set gap, for optical printing a 60 nm Au NP next to a ND with a height of 50 nm and a diameter of 70 nm. Printing was possible at every set gap, and gaps below 100 nm were achieved. The inset shows a dark-field image of fabricated ND-NP dimers for a 0 nm set gap, together with a SEM image of an exemplary dimer. The experiments were repeated using larger disks with a height of 50 nm and diameters of 100 and 150 nm. Figure 6d compiles the results of all the experiments described so far, including printing close to NPs and NDs on glass and sapphire substrates. The printing error (calculated as the deviation between the set gap and the average experimental gap) is plotted versus the calculated temperature increase of the fixed NP or ND. Temperature estimations considered the absorption cross section, substrate, laser power, and distance between the particle and the printing beam. Notably, the temperature increase of the fixed particle is a good predictor of the error when printing a second particle, strongly indicating that the resolution of optical printing is limited by photothermal effects.

Several plausible mechanisms for the photothermal repulsion were evaluated by Gargiulo et al.[19], including natural convection, Marangoni convection, thermo-osmotic flows, and thermophoresis. They found that a thermophoretic interaction between NPs explains the experimental observations and that the other phenomena only play a minor or negligible role. Thermophoretic forces $F_{tph}$ were calculated based on temperature maps, using

$$F_{tph} = -6\pi\mu a D_T \nabla T$$

with µ the dynamic viscosity, $a$ the NP radius, $D_T$ the thermodiffusion coefficient and $\nabla T$ the temperature gradient. $D_T$ is typically positive. In that case, NPs move from hot to cold regions. Figure 6e schematically shows the forces active during optical printing of dimers. In addition to the optical force, thermophoretic repulsion pushes the second NP away from the hot fixed NP towards the cold solution. Figure 6f shows total force fields, including the optical and the thermophoretic force, acting on a 60 nm Au NP around a fixed Au NP. The position of the laser is marked with a vertical line. The three regimes, *interacting without printing, interacting with printing* and *non-interacting* are well predicted by the model. The upper panel corresponds to optical printing on glass at a set gap of 50 nm. At this condition, optical printing is impossible; every trajectory points away from the substrate. The middle panel corresponds to the same set gap of 50 nm on sapphire. In this case, printing is possible, but the trajectories lead to a final position on the substrate with a larger gap than the target. Finally, the lower panel corresponds to a set gap of 240 nm on glass, where trajectories are not disturbed from the set point.

In the same manuscript, Gargiulo *et al.* evaluated a substrate with further enhanced heat dissipation. They included a highly thermally conductive reduced graphene oxide (rGO) monolayer onto a sapphire substrate, as schematized in Figure 6g. Using this substrate, they demonstrated optical printing of dimers of connected Au NPs, as shown in the SEM images of the same figure. However, the properties of the rGO film were not uniform, preventing the realization of a more detailed study. Nonetheless, these experiments indicate that photothermal repulsions could be completely avoided if effective thermal management strategies are implemented.

*Challenge 3: Selectivity of NPs with optical printing*

It is often the case, especially for recently developed NPs, that colloidal suspensions comprise NPs spanning a wide range of sizes and even different shapes. Since the optical properties of NPs are usually strongly dependent on size and shape, optical printing can be used wisely to selectively separate and immobilize different NP species of a colloid.[36] For example, in 2016, Huergo *et al.*[14] used optical printing to investigate the synthesis of Au NPs by the reduction of $HAuCl_4$ with $Na_2S$, which resulted in a complex suspension including spherical NPs, equilateral triangular NPs, and triangular NPs with rounded corners (Figure 7a). Using on-resonance optical printing, they could sort NPs based on their plasmon resonance. They evaluated the synthesis product for two different reaction times by optical printing the resulting colloid with three different lasers: 532 nm, 808 nm, and 1064 nm (Figure 7b). Dark-field images of the NPs printed with 808 nm and 532 nm (Figure 7c) revealed NPs with different dispersion spectra depending on the laser used to print them. SEM inspection confirmed the shape selectivity of optical printing at the three wavelengths (Figure 7d). Rather equilateral, sharp-tipped triangular NPs were selectively printed with the 1064 nm laser. Using 808 nm laser, triangular NPs with truncated edges were printed. Finally, the 532 nm laser printed preferentially spherical NPs. Figure 7e shows example scattering spectra of one of the spherical NPs printed at 532 nm and one triangular NP with truncated tips printed at 808 nm. The selective on-resonance optical printing of these two types of NP is clearly explained by their distinct plasmon resonances at around 530 nm and 780 nm.

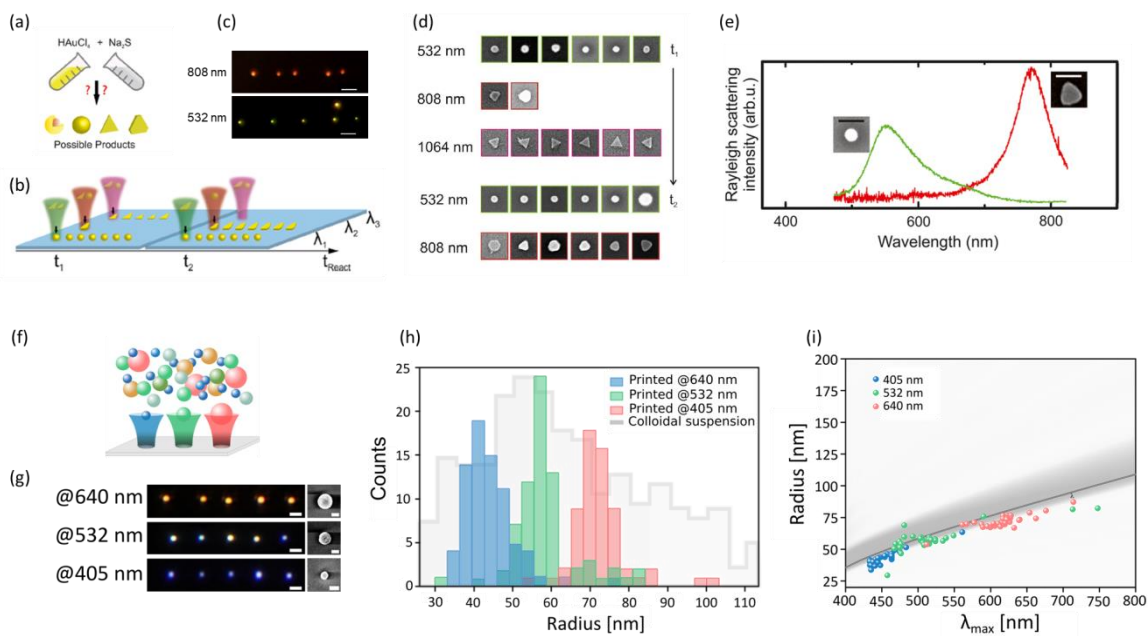

**Figure 7.** (a-e) Selective optical printing of Au NPs produced by reducing HAuCl$_4$ with Na$_2$S. (a) Scheme of the synthesis and products. (b) Scheme of the selective optical printing. (c) Dark-field images of NPs optically printed from the resulting colloid with a laser of 808 nm (top) and 532 nm (bottom). Scale bar: 20 µm. (d) SEM images of NPs obtained at two different reaction times and printed with different wavelengths. (e) Single particle scattering spectra of printed NPs using the 808 nm laser (green line) and the 1064 nm laser (red line). Scale bar: 100 nm. (f-i) Size-selective optical printing of Si NPs. (f) Scheme of the size selective optical printing using different wavelengths. (g) Dark-field images of Si NPs optically printed from the same colloid with three different wavelengths; SEM images representing each size of the printed NP. Dark-field scale bar: 2 µm, SEM scale bar: 100 nm. (h) Size distributions of Si NPs optically printed at 405, 532, and 640 nm. In the background, the size distribution of the original colloidal suspension is shown at scale for comparison. (i) Radius versus scattering maximum of the magnetic resonance for 115 Si NPs printed with the three wavelengths. The gray solid line is the theoretical prediction $\lambda = 2Rn$. The gray area represents the calculated dependency of radiation pressure efficiency on the wavelength and the particle radius. (a-e) Adapted with permission from Ref. 14 Copyright 2016 American Chemical Society. (f-i) Adapted with permission from Ref. 13 Copyright 2019 American Chemical Society.

Apart from plasmonic NPs, dielectric NPs, particularly silicon (Si) NPs, have gained considerable interest due to their size-dependent strong magnetic dipolar resonances in the visible range[37] and lower losses compared to their metallic counterparts.[38] However, their widespread application has been limited because their preparation on monodisperse colloids remains challenging. In 2019, Zaza *et al.* studied the optical printing of Si NPs as well as their sorting[13] by size, which had been proposed theoretically a few years before.[39] In this work, Si NPs were optically printed from a polydisperse colloidal suspension using three different wavelengths (405 nm, 532 nm, and 640 nm). Figure 7g shows dark-field images of the Si NPs printed at each wavelength, along with a representative SEM image. Clearly, the Si NPs can be printed in a size selective manner by tuning the laser wavelength. The size selectivity is better observed in the size distributions shown in Figure 7h. In grey, the size distribution of the initial colloid is shown, which includes NPs with radii ranging from 20 nm to 110 nm. Optical printing selects subpopulations of NPs with narrow size distributions with widths of around 10 nm. Individual Si NPs optically printed at the different wavelengths were subjected to correlated single particle scattering spectroscopy and FE-SEM imaging. Figure 7i shows the NP radius ($R$) versus the position of the scattering maximum ($\lambda_{max}$) for 115 optically printed Si NPs. The data follows very well the theoretical prediction for the magnetic dipolar resonance $\lambda_{max} = 2Rn$ (with $n$ the NP refractive index). The presence of a unique magnetic resonance that predominantly contributes to the optical force for a particular size enables the size-selective optical printing. For a given wavelength, differential optical forces can be applied to the NPs, so that only NPs of a certain size would be capable of surpassing the electrostatic barrier and be printed. It is important to point out that laser power is also a key parameter. If the laser power is too high, the size selectivity is reduced and eventually lost, as shown in Figure S4 of the ESI.

These two studies clearly show that on-resonance optical printing can be used to print selected subpopulations of NPs from a mixed colloid, according to their differential interaction with light.

*Challenge 4: Photostability during optical printing*

Optical printing was developed with the perspective of utilizing virtually any colloidal NP on solid substrates, circuits, and devices. Achieving that goal would imply that the optical printing process should not alter the properties, morphology, or functionalities of the NPs. However, the high laser power densities employed during the printing impose a challenge on this respect because they could lead to NP deformation or the degradation of surface functionalities. The dark-field images shown in Figure 8a illustrate the harmful effects of high laser power density when printing spherical Ag NPs on-resonance. Most NPs printed at low laser power appear blue, as expected from their plasmon resonance. By contrast, at high laser power, NPs of various colors are printed. While this is partly due to the loss of size selectivity, the variations observed are larger than what is expected from the size distribution of the NPs in the original colloid, indicating that morphological or compositional changes have occurred. One reason for this could be the photo-oxidation of Ag.[40]

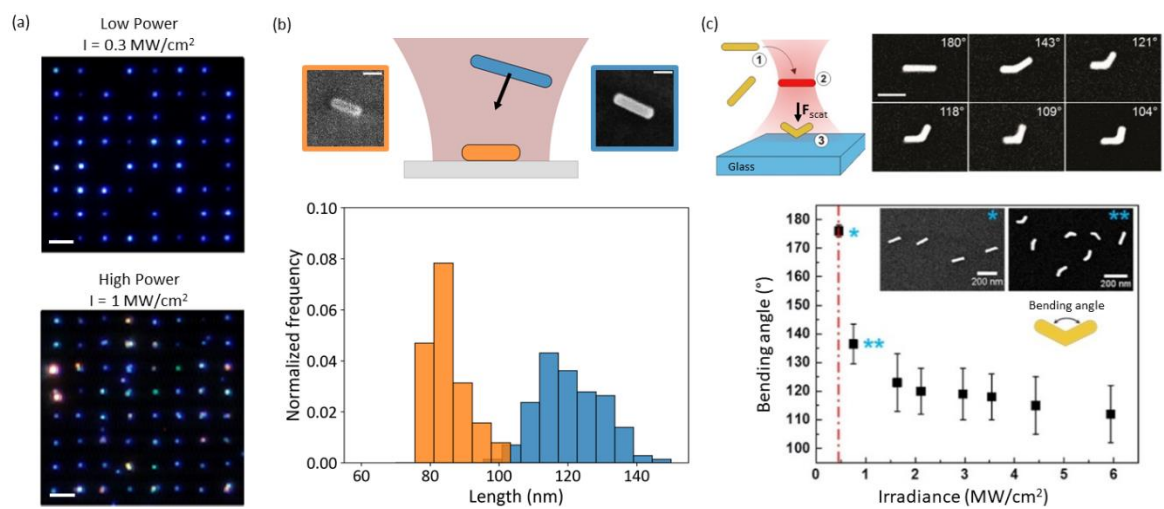

**Figure 8.** (a) Dark-field images of grids of 60 nm Ag NPs fabricated by on-resonance optical printing with a 405 nm laser at low (0.3 MW/cm$^2$) and (1 MW/cm$^2$) high irradiance conditions. Scale bars: 3 µm. (b, c) Reshaping of Au NRs during optical printing. (b) Histogram of length between before (blue) and after (orange) printing. Dimensions were acquired by FE-SEM. Scale bars: 50 nm. (c) Au NR bending during printing. The bending angle depends on the irradiance as confirmed by SEM images. Scale bar: 100 nm. (a) Adapted with permission from Gargiulo *et al.*[18] Copyright 2017 American Chemical Society. (c) Reproduced with permission from Ref. [25]. Copyright 2016 American Chemical Society.

Anisotropic NPs can undergo morphological changes during optical printing. For example, our group studied the optical printing of Au NRs with an aspect ratio of ~4.4 (longitudinal plasmon resonance at 896 nm) on glass substrates at wavelength of 808 nm, and noticed a length reduction upon printing. Figure 8b shows two FE-SEM images of representative NRs before (blue) and after (orange) printing, as well as length distributions of Au NRs before and after the printing process. The length distribution of printed Au NRs is centered at shorter lengths and with practically no overlap with the length distribution of Au NRs before printing. Thus, it can be concluded that the shorter length of the printed NRs is not due to selective printing, and that reshaping of the NRs is occurring during the printing process. The exact mechanism of this reshaping is yet unclear. Considering the experimental parameters, the temperature of the Au NRs may be increased up 535 K during printing (see Section S5 of the ESI for more details). Although this value is far below the bulk melting temperature of gold, reshaping can still occur, for example due to activated diffusion of surface atoms.[41]

The light-induced morphological changes during printing can potentially be used as a tool to control the shape of the printed nanostructure. In 2016, Babynina *et al.* demonstrated that it is possible to bend Au NRs (aspect ratio 5.9, longitudinal plasmon resonance at 1064 nm) when printing on-resonance.[25] Furthermore, the bending was more pronounced for higher printing irradiances, as illustrated in Figure 8c. These authors speculated that the bending occurs due to the combination of plasmonic heating with high hydrodynamic pressure working on the tips of the NR as it moves through water towards the substrate, pushed by the optical force. Furthermore, they observed the presence of twin grain boundaries near the bent nanorod vertex, indicating the reorganization of the gold atoms.

Heterogeneous or hybrid NPs are also prone to reshaping during the printing process. For example, the optical printing of silica-gold core-shell NPs ($SiO_2$@Au) was investigated in our group. $SiO_2$@Au NPs (150 nm $SiO_2$ core - 20 nm Au shell) were optically printed on a glass substrate using a 640 nm laser beam. The morphology and composition of the printed NPs was not always preserved. Three outcomes were observed, as depicted in Figure 9a: i) Properly printed, apparently intact $SiO_2$@Au core-shell NPs, ii) dimers of $SiO_2$ and Au NPs, and iii) $SiO_2$ NPs alone. Figure 9b shows a dark-field image with an exemplary grid of optically printed $SiO_2$@Au NPs. A FE-SEM image of the area in the white box is shown in Figure 9c, with examples of the three possible outcomes. The NP with a light-blue box has a diameter of (191 ± 4) nm, compatible with a 150@20 nm

SiO$_2$@Au core-shell NP. Furthermore, Figure 9d (blue solid line) shows the scattering spectrum of that individual NP (see more statistics in Section S6 of the ESI), nicely overlapping with the analytical spectrum calculated using Mie theory (dashed black line). On the other hand, a dimer can be seen in the orange box. The left particle in the dimer has a diameter of (142 ± 4) nm, potentially corresponding to the SiO$_2$ core. The right NP has a diameter of (151 ± 4) nm, roughly matching the volume of the average gold shell in the SiO$_2$@Au NPs. The scattering spectrum of this dimer is shown in Figure 9d, orange solid line. Interestingly, it matches the scattering spectra of a Au sphere with a diameter of 152 nm, indicating that the particle in the right of the dimer is a Au NP. The SiO$_2$ NP has a weak scattering cross section and is not visible by dark-field microscopy or spectroscopy. This is the case for the left particle in the dimer and the NP in the green box. These observations indicate that during the printing process, the energy absorbed by the NP could induce adatom diffusion of the gold shell, producing a gold NP, which may be lost during the printing process, or end up attached to the silica core forming silica-Au NP dimer.

SiO$_2$@Au NPs exhibit broad optical resonances over a large region of the visible and IR spectrum, as shown in the single particle scattering spectra of Figure 9d, meaning that they can be printed with many different wavelengths. In such cases, a strategy to minimize photothermal damage of NPs during the printing is to find a wavelength that maximizes the axial optical force while minimizing the absorption. The ratio between the radiation pressure efficiency $Q_{pr}$ and the absorption efficiency $Q_{abs}$ as a function of the wavelength, for the SiO$_2$@Au NPs, is shown in Figure 9e (black curve).[42] Interestingly, the efficiency ratio $Q_{pr}/Q_{abs}$ has a value of 1.5 at 640 nm, but peaks at 8.3 at 900 nm. A detailed analysis of $Q_{pr}$ and $Q_{abs}$ indicates that, in comparison to 640 nm, a printing beam of 900 nm would double the optical force while reducing the photogenerated heat by a factor of 2.8 (Figure S7). Therefore, printing with 900 nm would reduce photothermal damage significantly and could be the key to print intact SiO$_2$@Au NPs. On the contrary, considering that optical printing of bare SiO$_2$ NPs is challenging due to their weak optical response, using a wavelength that maximizes $Q_{pr}/Q_{abs}$, e.g. around 500 nm, could enable the printing of bare SiO$_2$ NPs using the SiO$_2$@Au NPs as precursors. It is interesting to remark that this strategy is not always available. For instance, the $Q_{pr}/Q_{abs}$ vs. wavelength for 60 nm spherical Au NP (Figure 9e, red curve) presents only minor variations.

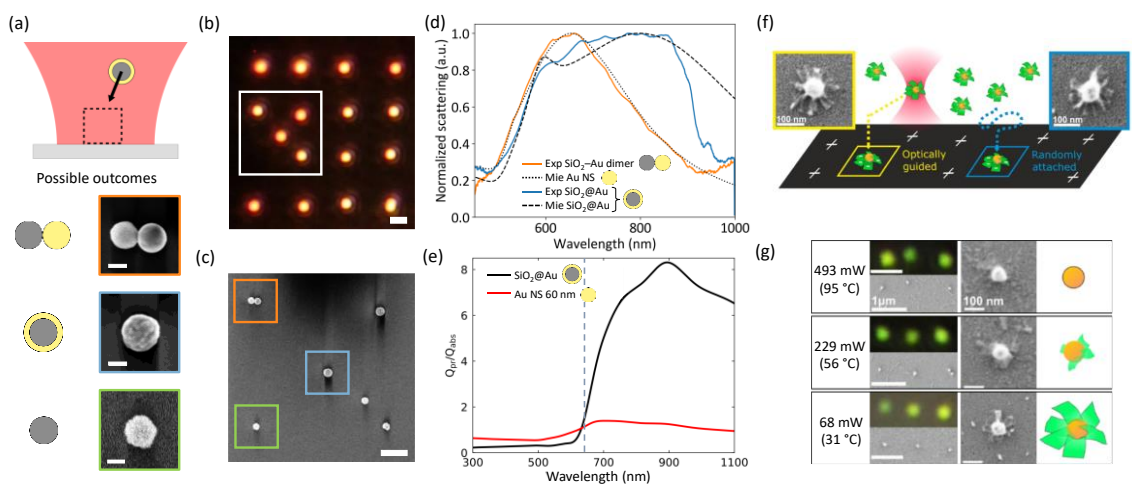

**Figure 9.** (a) Outcomes of the optical printing of $SiO_2$@Au core-shell NPs. From top to bottom, sketch and SEM images of Au-$SiO_2$ dimer, $SiO_2$@Au core-shell NP, and $SiO_2$ core. Diameters of Au and $SiO_2$ NPs of the dimer are (151 ± 4) nm and (142 ± 4), respectively. The diameter of the $SiO_2$@Au core-shell NP is (191 ± 4) nm. The diameter of the $SiO_2$ core is (140 ± 4) nm. Scale bar = 100 nm in all images. (b) Dark-field image of an array of optically printed $SiO_2$@Au NPs on a glass substrate, under water. Scale bar = 1 µm. (c) SEM image of the region highlighted in (d, white square). A Au-$SiO_2$ dimer (orange), a core-shell NP (blue) and a silica core (green) are outlined. Scale bar = 500 nm. (d) Measured and calculated (Mie) scattering spectra of a spherical gold NP (152 nm) and a $SiO_2$@Au core-shell (150 nm $SiO_2$ core - 20 nm Au shell) under water. (e) Ratio between the radiation pressure efficiency $Q_{pr}$ and the absorption efficiency $Q_{abs}$ as a function of the wavelength calculated for a $SiO_2$@Au core-shell NP (150 nm $SiO_2$ core - 20 nm Au shell; black line) and a 60 nm spherical Au NP (red line). (f) Sketch of optical trapping and printing of Au@DNA-origami hybrids. A CW laser at 1064 nm is used to trap the nanohybrids. The structures are then directed to the surface of a positively PDDA charge glass substrate and electrostatically fixed there. (g) Effect of laser trapping power on the stability of Au@DNA-origami hybrids. For each laser power (1064 nm, FWHM = 1 µm), the predicted temperatures of the Au NPs are indicated, example SEM and dark-field images, and a sketch of the nanohybrid after printing are shown. (g-h) Adapted with permission from Ref. [43]. Copyright 2012 American Chemical Society.

The preservation of the surface functionality of colloidal NPs is another key challenge. During the printing process, the temperature of NPs can increase several hundreds of degrees, destabilizing their surface functionalization. For example, due to their high affinity for gold [44], thiols are commonly employed to functionalize Au NPs with biological and organic molecules. While the Au-S bond breaks at 470 K[45], a 60 nm Au NP may reach temperatures of up to 645 K during on-resonance printing, as shown in Figure 6a. However, it should be noted that the high temperatures during the printing process are only reached during a short time of a few milliseconds or less,[18] so the stability of the functionalizing groups will depend on the particular kinetics of each reaction.

The possibility of printing functionalized Au NP was experimentally tested by Do *et al.*[43] They printed 60 nm Au NPs surface-functionalized with DNA-origami[46] sheets (hereafter Au@DNA-origami) as illustrated in Figure 9f. The Au NPs were functionalized with a thiolated chain of 24 thymine bases, to which the DNA sheets were attached by DNA hybridization. As DNA origami is also temperature sensitive,[47,48] they performed a so called *soft optical deposition* consisting of trapping the Au@DNA-origami NPs with a laser off-resonance, and subsequently approaching them to a positively charged substrate, *i.e.*, having the opposite charge than the negatively charged NPs. The elimination of the substrate repulsion allowed them to use sufficiently low laser powers to significantly decrease the absorption and managed to print intact Au@DNA-origami NPs (Figure 9g). However, this approach has the important disadvantage that NPs may attach to the substrate spontaneously in random positions due to the electrostatic attraction. The authors minimized this by using highly diluted colloids, which in turn made the printing process very slow. Increasing the trapping power to 229 mW, which corresponded to NP temperatures of 329 K, led to degradation of the 2D DNA sheets (Figure 9g). A later report by Goodman *et al.* [49] found similar results when illuminating $SiO_2$@Au NPs coated with double-stranded DNA. These results indicate that, when performing optical printing of DNA-modified NPs, it is more likely to reach the melting temperature of the DNA duplex than breaking the Au-S bond.[50]

The different examples presented in this section demonstrate that light absorption and heating during the printing process may alter the NP composition, shape, or surface functionalization. Nonetheless, there are clear avenues to optically print intact NPs, such as looking for wavelengths that maximize the ratio between optical forces and absorption, or finely tuning the NP-substrate repulsion to minimize the laser power required for printing. In addition, minimizing unnecessary exposure to the laser beam could be achieved by automating the printing process to rapidly turn off the laser upon printing of a NP.

*Conclusions and perspectives*

Optical printing is a nanofabrication technique that allows the positioning, orientation, combination, and selection of individual NPs from a colloidal suspension into arbitrary patterns on a variety of substrates. Here, we have reviewed the advances and contributions to the field since its introduction one decade ago and discussed them in the context of four challenges: accuracy, resolution, selectivity, and photostability.

During the last years, a detailed description of the printing mechanism was built, and the most relevant physical parameters were identified. The positional accuracy of optical printing is determined by the balance between the DLVO repulsion and the axial optical force at the substrate, and by the shape of the optical force field at the printing wavelength. For Au and Ag NPs, using lasers tuned to the plasmon resonance of each type of NP and employing low powers led to higher positional accuracy, in the range of 50 to 150 nm.

The resolution of optical printing is compromised by repulsive thermophoretic forces that arise from absorption and heating by the fixed NPs. It must be noted that this fact, usually discussed as a limitation, is beneficial to print arrays of isolated single NPs. On the other hand, if the goal is to connect NPs, different strategies can be used. One approach is to reduce the optical interaction between the already printed NPs and the printing wavelength. This strategy was proven useful to connect two distinct NPs. However, it has limited capabilities for the fabrication of dimers consisting of alike NPs or arrays of three or more connected NPs. The second approach is to implement heat management strategies to reduce the temperature increase during printing. In specific cases, such as using a sapphire substrate covered with a reduced graphene oxide layer, this strategy totally suppresses the thermophoretic repulsion. Further research in this direction may lead to practical configurations of more general use. A third, rather unexplored strategy is to reduce thermophoretic interactions by lowering the magnitude (or inverting the sign) of the thermo-diffusion coefficient $D_T$. Even though important advances on the understanding of thermophoretic interactions have been recently achieved,[51] a clear insight regarding heat conductive NPs is still missing, and further studies would be required in that direction.

First studies of optical printing were made with spherical metallic (Au and Ag) NPs. More recently, optical printing was applied to differently shaped Au NPs, multi-component inorganic NPs, inorganic-organic hybrid nanosystems, and even 2d materials.[52] In these cases, ensuring the stability of the NPs during the printing process is the main challenge. Here, a few guidelines were summarized to tackle this, such as finding printing wavelengths that maximize optical forces while minimize the generated heat and reducing the electrostatic barrier with the substrate. In the future, we envision the optical printing of plasmonic and hybrid nanoparticles with a variety of morphologies and surface supramolecular modifications, by correctly selecting the optical printing conditions of these materials. When monodisperse colloids are not available, on-resonance optical printing can be used to selectively print subpopulations of NPs, as long as they present differentiable spectral responses. In this case, calculations of the optical force are particularly useful to predict the printing selectivity as a function of the wavelength.

Most demonstrations of optical printing used glass substrates. However, optical printing can also incorporate colloidal NPs into specific locations of more complex substrates. It has already been shown that Au NPs can be printed into photonic crystals,[53] living cells,[54] or 2D carbon materials.[19] We envision further integration of plasmonic NPs into photonic systems such as 2D planar waveguides,[55] hollow-core 3D waveguides,[56] microstructures,[57] optical fibers,[58] or microcavities.[59]

The progress reviewed here expands the field of application of optical printing as a template-free and versatile nanofabrication strategy for positioning colloidal NPs on a solid surface. The insights into the physical mechanism of optical printing provide clear guidelines for devising new, optimized optical printing strategies with high accuracy, resolution, selectivity, and photostability. We envision the fabrication of complex functional nanostructures and circuits based on connected NPs, and their hierarchical integration to on-chip photonic systems.

Finally, optical printing can accelerate new research too. Having an ordered array of equivalent NPs facilitates the implementation of automated characterization routines and the accumulation of larger datasets. For example, it has been used to study plasmon-assisted chemical reactions on individual NPs,[20] and the photothermal response of single Au NPs on different substrates.[27] The detailed knowledge about the mechanism of optical printing enables its use to investigate other phenomena at the nanoscale, such as thermophoresis and heat dissipation.

SUPPLEMENTARY MATERIAL

See supplementary material for more examples on the aging of arrays of NPs prepared by optical printing, details on the calculations of optical forces, the power dependance of selectivity in the printing of Si NPs, details on the Au NR reshaping, and different extinction spectra.

ACKNOWLEDGEMENTS

F.D.S. acknowledges the support of the Alexander von Humboldt Foundation. J.G. acknowledges the PRIME programme of the German Academic Exchange Service (DAAD) with funds from the German Federal Ministry of Education and Research (BMBF) and the Royal Society of Chemistry for the Research Fund R20-7244. We thank Hilario Boggiano for assistance with scanning electron microscopy measurements.

FUNDING INFORMATION

This work has been funded by CONICET, ANPCYT Project PICT-2017-0970.

REFERENCES

[1] H. Zhang, C. Kinnear, and P. Mulvaney, Adv. Mater. **n/a**, 1904551 (2019).

[2] N. Vogel, M. Retsch, C.A. Fustin, A. Del Campo, and U. Jonas, Chem. Rev. **115**, 6265 (2015).

[3] J.B. Lee, H. Walker, Y. Li, T.W. Nam, A. Rakovich, R. Sapienza, Y.S. Jung, Y.S. Nam, S.A. Maier, and E. Cortés, ACS Nano **14**, 17693 (2020).

[4] V. Flauraud, M. Mastrangeli, G.D. Bernasconi, J. Butet, D.T.L. Alexander, E. Shahrabi, O.J.F. Martin, and J. Brugger, Nat. Nanotechnol. **12**, 73 (2016).

[5] P.Y. Chiou, A.T. Ohta, and M.C. Wu, Nature **436**, 370 (2005).

[6] A. Jamshidi, P.J. Pauzauskie, P.J. Schuck, A.T. Ohta, P.Y. Chiou, J. Chou, P. Yang, and M.C. Wu, Nat. Photonics **2**, 86 (2008).

[7] L. Lin, X. Peng, and Y. Zheng, Chem. Commun. **53**, 7357 (2017).

[8] L. Lin, X. Peng, Z. Mao, W. Li, M.N. Yogeesh, B.B. Rajeeva, E.P. Perillo, A.K. Dunn, D. Akinwande, and Y. Zheng, Nano Lett. **16**, 701 (2016).

[9] B.B. Rajeeva, M.A. Alabandi, L. Lin, E.P. Perillo, A.K. Dunn, and Y. Zheng, J. Mater. Chem. C **5**, 5693 (2017).

[10] O.M. Maragò, P.H. Jones, P.G. Gucciardi, G. Volpe, and A.C. Ferrari, Nat. Nanotechnol. **8**, 807 (2013).

[11] A. Lehmuskero, P. Johansson, H. Rubinsztein-Dunlop, L. Tong, and M. Käll, ACS Nano **9**, 3453 (2015).

[12] A.I. Kuznetsov, A.E. Miroshnichenko, M.L. Brongersma, Y.S. Kivshar, and B. Luk'yanchuk, Science (80-. ). **354**, (2016).

[13] C. Zaza, I.L. Violi, J. Gargiulo, G. Chiarelli, L. Schumacher, J. Jakobi, J. Olmos-Trigo, E. Cortes, M. König, S. Barcikowski, S. Schlücker, J.J. Sáenz, S.A. Maier, and F.D. Stefani, ACS Photonics **6**, 815 (2019).

[14] M.A. Huergo, C.M. Maier, M.F. Castez, C. Vericat, S. Nedev, R.C. Salvarezza, A.S. Urban, and J. Feldmann, ACS Nano **10**, 3614 (2016).

[15] M.J. Guffey and N.F. Scherer, Nano Lett. **10**, 4032 (2010).

[16] A.S. Urban, A. a. Lutich, F.D. Stefani, and J. Feldmann, Nano Lett. **10**, 4794 (2010).

[17] J. Gargiulo, S. Cerrota, E. Cortés, I.L. Violi, and F.D. Stefani, Nano Lett. **16**, 1224 (2016).

[18] J. Gargiulo, I.L. Violi, S. Cerrota, L. Chvátal, E. Cortés, E.M. Perassi, F. Diaz, P. Zemánek, and F.D. Stefani, ACS Nano **11**, 9678 (2017).

[19] J. Gargiulo, T. Brick, I.L. Violi, F.C. Herrera, T. Shibanuma, P. Albella, F.G. Requejo, E. Cortés, S.A. Maier, and F.D. Stefani, Nano Lett. **17**, 5747 (2017).

[20] I.L. Violi, J. Gargiulo, C. Von Bilderling, E. Cortés, and F.D. Stefani, Nano Lett. **16**, 6529 (2016).

[21] J. Do, M. Fedoruk, F. Jäckel, and J. Feldmann, Nano Lett. **13**, 4164 (2013).

[22] A. Königer and W. Köhler, ACS Nano **6**, 4400 (2012).

[23] A. Foti, C. D'Andrea, V. Villari, N. Micali, M.G. Donato, B. Fazio, O.M. Maragò, R. Gillibert, M.L. de la Chapelle, and P.G. Gucciardi, Materials (Basel). **11**, (2018).

[24] S. Bernatová, M.G. Donato, J. Ježek, Z. Pilát, O. Samek, A. Magazzù, O.M. Maragò, P. Zemánek, and P.G. Gucciardi, J. Phys. Chem. C **123**, 5608 (2019).

[25] A. Babynina, M. Fedoruk, P. Kühler, A. Meledin, M. Döblinger, and T. Lohmüller, Nano Lett. **16**, 6485 (2016).

[26] J.N. Israelachvili, *Intermolecular and Surface Forces* (Academic Press, London, 1992).

[27] M. Barella, I.L. Violi, J. Gargiulo, L.P. Martinez, F. Goschin, V. Guglielmotti, D. Pallarola, S. Schlücker, M. Pilo-Pais, G.P. Acuna, S.A. Maier, E. Cortés, F.D. Stefani, E. Cortes, F.D. Stefani, M. Barella, I.L. Violi, L.P. Martinez, F. Goschin, V. Guglielmotti, D. Pallarola, S. Schlucker, M. Pilo-Pais, G.P. Acuna, and S.A. Maier, ACS Nano **15**, 2458 (2021).

[28] Y. Bao, Z. Yan, and N.F. Scherer, J. Phys. Chem. C **118**, 19315 (2014).

[29] S. Nedev, A.S. Urban, A. a Lutich, and J. Feldmann, Nano Lett. **11**, 5066 (2011).

[30] L. Ling, H.L. Guo, X.L. Zhong, L. Huang, J.F. Li, L. Gan, and Z.Y. Li, Nanotechnology **23**, (2012).

[31] A.S. Urban, M. Fedoruk, S. Nedev, A. Lutich, T. Lohmueller, and J. Feldmann, Adv. Opt. Mater. **1**, 123 (2013).

[32] A.S. Urban, S. Carretero-Palacios, A.A. Lutich, T. Lohmüller, J. Feldmann, and F. Jäckel, Nanoscale **6**, 4458 (2014).

[33] K. Dholakia and P. Zemánek, Rev. Mod. Phys. **82**, 1767 (2010).

[34] Z. Yan, S.K. Gray, and N.F. Scherer, Nat. Commun. **5**, 3751 (2014).

[35] L. Chvátal, O. Brzobohatý, and P. Zemánek, Opt. Rev. **22**, 157 (2015).

[36] M. Ploschner, T. Čižmár, M. Mazilu, A. Di Falco, and K. Dholakia, Nano Lett. **12**, 1923 (2012).

[37] A.I. Kuznetsov, A.E. Miroshnichenko, Y.H. Fu, J. Zhang, and B. Lukyanchukl, Sci. Rep. **2**, 1 (2012).

[38] P. Albella, M.A. Poyli, M.K. Schmidt, S.A. Maier, F. Moreno, J.J. Sáenz, and J. Aizpurua, J. Phys. Chem. C **117**, 13573 (2013).

[39] D.A. Shilkin, E. V. Lyubin, M.R. Shcherbakov, M. Lapine, and A.A. Fedyanin, ACS Photonics **4**, 2312 (2017).

[40] N. Grillet, D. Manchon, E. Cottancin, F. Bertorelle, C. Bonnet, M. Broyer, J. Lermé, and M. Pellarin, J. Phys. Chem. C **117**, 2274 (2013).

[41] A.B. Taylor, A.M. Siddiquee, and J.W.M. Chon, ACS Nano **8**, 12071 (2014).

[42] J. Olmos-Trigo, M. Meléndez, R. Delgado-Buscalioni, and J.J. Sáenz, Opt. Express **27**, 16384 (2019).

[43] J. Do, R. Schreiber, A.A. Lutich, T. Liedl, J. Rodríguez-Fernández, and J. Feldmann,

Nano Lett. **12**, 5008 (2012).

[44] J.C. Azcárate, G. Corthey, E. Pensa, C. Vericat, M.H. Fonticelli, R.C. Salvarezza, and P. Carro, J. Phys. Chem. Lett. **4**, 3127 (2013).

[45] L. Poon, W. Zandberg, D. Hsiao, Z. Erno, D. Sen, B.D. Gates, and N.R. Branda, ACS Nano **4**, 6395 (2010).

[46] P.W.K. Rothemund, Nature **440**, 297 (2006).

[47] R. Huschka, J. Zuloaga, M.W. Knight, L. V. Brown, P. Nordlander, and N.J. Halas, J. Am. Chem. Soc. **133**, 12247 (2011).

[48] J. Stehr, C. Hrelescu, R.A. Sperling, G. Raschke, M. Wunderlich, A. Nichtl, D. Heindl, K. Kürzinger, W.J. Parak, T.A. Klar, and S.J. Feldmann, Nano Lett. **8**, 619 (2008).

[49] A.M. Goodman, N.J. Hogan, S. Gottheim, C. Li, S.E. Clare, and N.J. Halas, ACS Nano **11**, 171 (2017).

[50] F. Thibaudau, J. Phys. Chem. Lett. **3**, 902 (2012).

[51] S. Liu, L. Lin, and H.B. Sun, ACS Nano **15**, 5925 (2021).

[52] M.G. Donato, E. Messina, A. Foti, T.J. Smart, P.H. Jones, M.A. Iatì, R. Saija, P.G. Gucciardi, and O.M. Maragò, Nanoscale **10**, 1245 (2018).

[53] J. Do, K.N. Sediq, K. Deasy, D.M. Coles, J. Rodríguez-Fernández, J. Feldmann, and D.G. Lidzey, Adv. Opt. Mater. **1**, 946 (2013).

[54] M. Li, T. Lohmüller, and J. Feldmann, Nano Lett. 141211060551002 (2014).

[55] F. Peyskens, P. Wuytens, A. Raza, P. Van Dorpe, and R. Baets, Nanophotonics **7**, 1299 (2018).

[56] J. Kim, B. Jang, J. Gargiulo, J. Bürger, J. Zhao, S. Upendar, T. Weiss, S.A. Maier, and M.A. Schmidt, Anal. Chem. **93**, 752 (2021).

[57] M.G. Donato, V.P. Rajamanickam, A. Foti, P.G. Gucciardi, C. Liberale, and O.M. Maragò, Opt. Lett. **43**, 5170 (2018).

[58] G.Q. Ngo, A. George, R.T.K. Schock, A. Tuniz, E. Najafidehaghani, Z. Gan, N.C. Geib, T. Bucher, H. Knopf, S. Saravi, C. Neumann, T. Lühder, E.P. Schartner, S.C. Warren-Smith, H. Ebendorff-Heidepriem, T. Pertsch, M.A. Schmidt, A. Turchanin, and F. Eilenberger, Adv. Mater. **32**, 2003826 (2020).

[59] M.A. Schmidt, D.Y. Lei, L. Wondraczek, V. Nazabal, and S.A. Maier, Nat. Commun. **3**, 1108 (2012).